\documentclass[fleqn,12pt]{wlscirep}
\usepackage[utf8]{inputenc}
\usepackage[T1]{fontenc}
\usepackage{lineno}

\title{Visible to Ultraviolet Frequency Comb Generation in Lithium Niobate Nanophotonic Waveguides}

\author[1,2]{Tsung-Han Wu}
\author[3,4]{Luis Ledezma}
\author[1,2]{Connor Fredrick}
\author[1,2]{Pooja Sekhar}
\author[3]{Ryoto Sekine}
\author[3]{Qiushi Guo}
\author[4]{Ryan M. Briggs}
\author[3]{Alireza Marandi}
\author[1,2,5,*]{Scott A. Diddams}
\affil[1] {Time \& Frequency Division, National Institute of Standards and Technology, Boulder, CO 80305 USA}
\affil[2] {Department of Physics, University of Colorado, Boulder, CO 80309 USA}
\affil[3] {Department of Electrical Engineering, California Institute of Technology, Pasadena, California 91125, USA}
\affil[4] {Jet Propulsion Laboratory, California Institute of Technology, Pasadena, California 91109, USA}
\affil[5] {Electrical Computer \& Energy Engineering, University of Colorado, Boulder, CO 80309 USA}

\affil[*]{corresponding author: scott.diddams@colorado.edu}

\begin{abstract}
The introduction of nonlinear nanophotonic devices to the field of optical frequency comb metrology has enabled new opportunities for low-power and chip-integrated clocks, high-precision frequency synthesis, and broad bandwidth spectroscopy.  However, most of these advances remain constrained to the near-infrared region of the spectrum, which has restricted the integration of frequency combs with numerous quantum and atomic systems in the ultraviolet and visible. Here, we overcome this shortcoming with the introduction of multi-segment nanophotonic thin-film lithium niobate (LN) waveguides that combine engineered dispersion and chirped quasi-phase matching for efficient supercontinuum generation via the combination of $\chi^{(2)}$ and $\chi^{(3)}$ nonlinearities. With only 90 pJ of pulse energy at 1550 nm, we achieve gap-free frequency comb coverage  spanning 330 to 2400 nm. The conversion efficiency from the near-infrared pump to the UV-Visible region of 350-550 nm is nearly 20\%. Harmonic generation via the $\chi^{(2)}$ nonlinearity in the same waveguide directly yields the carrier-envelope offset frequency and a means to verify the comb coherence at wavelengths as short as 350 nm. Our results provide an integrated photonics approach to create visible and UV frequency combs that will impact precision spectroscopy, quantum information processing, and optical clock applications in this important spectral window.

\end{abstract}
\begin{document}

\flushbottom
\maketitle

\thispagestyle{empty}


\section*{Introduction}
Optical frequency combs offer a powerful tool for precise spectroscopic measurements using a broad array of equidistant and mutually coherent comb elements whose frequencies can be tied to absolute standards. In the two decades since the first near infrared frequency combs were demonstrated, a tremendous range of scientific discovery and novel applications have been explored across the optical spectrum--ranging from the terahertz (THz) to the ultraviolet (UV)\cite{Scott2020,Fortier2019}. Still, efficient and reliable frequency comb coverage across the full visible and UV has been challenging, and is limiting multiple impactful research topics.  For example, absolute UV frequency knowledge is critical to efficiently reach and reference the cooling and trapping transitions of atoms and ions used in optical clocks, quantum computing and sensing  \cite{ludlow2015,HAFFNER2008}. In addition, frequency combs with 10-30 GHz mode spacing play an important role in astronomical spectrograph calibration for exoplanet science.  However, among all the work in this field, an Earth-Sun analog has still not been discovered. Doing so will require precision astronomical radial velocity measurements with $\leq10$ cm/s precision over multiple years, which includes robust 10+ GHz frequency combs across the peak of the Solar spectrum from 400 nm to 650 nm\cite{Fischer2016}. Finally, a broadband UV comb is central to high-precision spectroscopy aimed at characterizing and quantifying UV transitions in atmospheric gases and aerosols that are important climate science \cite{rs12203444}. 

The lack of suitable laser gain media in UV spectral region elevates the importance of nonlinear optical techniques to access this region. Up-conversion using harmonics generation with noble gases\cite{Kandula2010,Pupeza2021,gohle2005frequency,PhysRevLett.94.193201} and bulk crystals\cite{Yang2012,Lesko2021,Lesko2022}, or supercontinuum broadening with waveguides\cite{Iwakuni2016,YoonOh2017,Herr:18,Rutledge2021,Escale2020,Liu2019,Nakamura2021} or  optical fiber\cite{Mridha:16} are the most common methods to produce coherent UV frequency combs. However, those approaches have only partial spectral coverage or require intense pump pulses or cavity enhancement\cite{Peters:09,Pupeza2021,Pinkert:11}. This increases complexity and noise through high power amplification and makes it unfeasible to realize UV coverage with gigahertz-rate frequency comb sources. Additionally, it has been observed that the ultraviolet light damages commonly used silica based fiber and waveguides, such that  performance degrades on the timescale of hours to days\cite{YoonOh2017}. 

Here, we introduce a novel engineered combination of nonlinear optical advances that provide efficient and UV to visible gap-free frequency comb coverage. Instead of using the  $\chi^{(3)}$ nonlinear process in the anomalous dispersion regime, we leverage domain engineering and the strong $\chi^{(2)}$ nonlinearity in the normal dispersion regime of the lithium niobate nanophotonics platform. Lithium niobate (LN) has a long history as one of the most widely used nonlinear optical materials, and recent developments in nano-scale fabrication open new opportunities to engineer waveguides for integrated supercontinuum and frequency comb generation at picojoule (pJ) and lower pulse energies \cite{Honardoost2020,Jankowski2020,Okawachi2020,Escale2020,Roy2022}.  In recent work, techniques were introduced for low energy measurement of the carrier-envelope offset frequency that take advantage of the simultaneous $\chi^{(2)}$ and $\chi^{(3)}$ nonlinearities\cite{Okawachi2020}. However, in these cases the continuous spectral bandwidth extends to approximately 600 nm, leaving at least 350 THz of the spectrum, at wavelengths in the UV, without spectral coverage.  Except for a few notable cases of narrowband phase-matched frequency conversion in material platforms like aluminum nitride\cite{Liu2019} and silica\cite{YoonOh2017}, the visible and near UV region of the spectrum has been largely inaccessible for frequency combs generated with nonlinear nanophotonics. 

Our work fills this significant spectral gap with continuous UV to visible frequency comb coverage. This is accomplished  with a novel multi-segment nonlinear waveguide design in thin-film lithium niobate by combining the engineering of dispersion and quasi-phase matching. The initial segment consists of a low dispersion waveguide that acts to spectrally broaden the 1550 nm pulses from a robust and technologically mature 100 MHz Er:fiber comb. The second segment includes  the engineering of quasi-phase matching which is poled with longitudinally chirped periods that enhance the generation of visible and near-UV via harmonic and cascaded $\chi^{(2)}$ processes. The combination of nanophotonic confinement and engineered dispersion and phase matching leads to an exceptional efficiency, where greater than 20\% of the <100 pJ 1550 nm pulse energy is spectrally translated to form a gap-free continuum in the spectral region of 350-550 nm. Low noise and optical coherence across 350 THz is verified by heterodyne measurements revealing kilohertz relative linewidths and a noise floor near the shot-noise limit. 

Our measurements are supported by numerical modeling with a single-envelope equation, yielding good agreement with experiments and the prediction of still greater increases in efficiency with optimized waveguide designs.  Simulations further support preliminary measurements with 1550 nm pulses at 10 GHz, illustrating the clear path to full visible-to-UV spectral coverage at 100x greater repetition rate with <100 pJ pulses. Together, our results highlight the means to achieve compact and robust frequency comb coverage across spectral bands critical for challenging and impactful applications in quantum sensing and computing, precision astronomical spectroscopic calibration, and broad bandwidth frequency comb spectroscopy. Furthermore, similar design strategies can be implemented for coherent frequency comb spectra across the full transparency window of lithium niobate, from 330 nm to 5 $\mu$m. 

\section*{Experiment}
Figure 1 shows the important details of our experiments and the nanophotonic LN waveguides we employ. Figure 1a depicts the waveguides that are etched from a 710 nm film of LN on oxide to be 350 nm high and 1800 nm wide, with a 1550 nm mode area of approximately 1 µm$^2$. The waveguide dispersion at 1550 nm is calculated to be slightly anomalous, and increases to a large normal value below 1000 nm (see Supplement). In the longitudinal direction, we introduce a segmented structure with 3 mm of unpoled region of the waveguide.  This is followed by 3.6 mm in which we implement a chirp in the poling period, $\Lambda$, that decreases linearly along the propagation direction from $\Lambda=$12.5 µm to $\Lambda=$2.5 µm. Instead of the more typical supercontinuum in anomalous dispersion, the 3 mm section employs $\chi^{(3)}$ self-phase modulation (SPM) to spectrally broaden the input pulse, while the chirped poling in the second section expands the near-infrared light to the visible and ultraviolet via $\chi^{(2)}$ and cascaded $\chi^{(2)}$ nonlinear processes. The evolution of the spectral broadening is discussed further below. 

The experimental setup is depicted in Figure 1b, including the pump laser source and diagnostics. We pump the waveguide with 1550 nm pulses from a 100-MHz Er:fiber frequency comb that is amplified to produce sub-100-fs pulses with 100 mW of average power. The power incident on the waveguide is varied without changing the pulse duration and a polarization maintaining (PM) lensed fiber is used for the input coupling.  We estimate coupling loss of 10 dB at each facet, such that 9 mW is coupled into the waveguide when pumped with 90 mW.  Under these conditions continuous spectral coverage extending from the near-infrared pump to the ultraviolet is achieved, as shown in Figure 1c. The extremely broad nature of the output spectrum requires that the output coupling be done with a reflective microscope objective that has low loss into the UV.  The inset of Figure 1c shows a photograph of the visible spectrum after dispersing the collimated output with a diffraction grating. 

The progression of the spectral expansion with input pulse energy is shown in Figure 2, where experimental and simulated spectra are presented side-by-side. The input pulse bandwidth is broadened from 30 nm to 400 nm in the first 3 mm of the unpoled waveguide (see Figure 1c).  In the next 3.6 mm, the $\chi^{(2)}$ nonlinearity generates second harmonic (2f, 384 THz, 780 nm), third harmonic (3f, 576 THz, 520 nm) and fourth harmonic (4f, 770 THz, 390 nm) with a driving pulse energy of less than 10 pJ.  As the pulse energy increases to tens of pJ, our simulations show that the spectral broadening is significantly enhanced in the poled region.  Notably, for pulse energies higher than 50 pJ, the spectral regions around 2f, 3f and 4f start to overlap, and with $\sim90$ pJ (9 mW on chip) the spectrum is continuous and gap free from 350 nm to 1000 nm. The progression is reflected in the simulations of Figure 2b, which employ a unidirectional propagation equation~\cite{Francois1991nonlinearFD,Kolesik2004UPE}.  Further details of the modeling are given in the Methods.  Over the full range of energies and spectral bandwidths, we see excellent agreement between the measurements and simulations in terms of the spectral coverage, spectral shape, and required pulse energy. As discussed below and in the Supplement, this level of agreement provides the basis for design improvements aimed at spectrally tailored frequency combs for optical clocks, astronomical spectrograph calibration, and spectroscopic sensing. 

Beside UV and visible comb generation at 100 MHz, we implement the same waveguide of Figure 1a to demonstrate visible comb generation when pumped with the output of a 10 GHz resonant EO frequency comb\cite{Sekhar2023}. These data are included as the dashed line in Figure 2a, which shows the spectrum of $2^{\rm nd}, 3^{\rm rd}$ and $4^{\rm th}$ harmonics with about 26 pJ of 50 fs pulses coupled into the waveguide. The plot is consistent with the 100 MHz data and simulations, with spectral bandwidth falling between the plots for 19 and 50 pJ. For this experiment, the UV intensity out of the waveguide was attenuated due to non-optimized output coupling into a multimode fiber of 1 m length. Even though the spectrum does not cover the full UV and visible range, the $4^{\rm th}$ harmonic power in the UV is still estimated to be on the order of 5-10 µW per comb mode. We see no damage to the nanophotonic waveguides, even with about 2.5 W of incident power at 10 GHz. With coupling losses reduced to 1-2 dB, as demonstrated by others\cite{He2019,Hu2021}, it should be possible to achieve continuous 10 GHz spectral coverage across the visible and UV with a little over 1 W of incident power.

The spectral overlap between the harmonics provides means to detect the carrier-envelope offset frequency, $f_{ceo}$ of the driving comb, as well as to verify the coherence of the comb structure.  This is accomplished by diffracting the output spectrum with a grating and detecting various regions of the spectrum with a Si avalanche photodiode (Si-APD, Figure 1b). Figure 3 shows $f_{ceo}$ detected at 620, 400 and 350 nm, respectively. At the longer wavelengths of 620 and 400 nm, the signal-to-noise ratio (SNR) is greater than 35 dB (10 kHz resolution bandwidth). And as shown for the 620 nm data, the noise floor is near the shot noise limit, indicating that higher powers or larger spectral bandwidths would lead to still higher SNR. The heterodyne beatnote in the ultraviolet (350 nm) provides the direct evidence of the coherence of the ultra-broad frequency comb generated with this unique chirped poling in thin-film LN waveguides. It is the presence of multiple frequency combs that leads to the observed $f_{ceo}$.  While for some applications this could introduce complications, these offsets could be easily eliminated by making $f_{ceo}=0$ for the entire comb \cite{Jones2000,Okubo2018}. We note that in spectroscopic situations, the presence of a secondary comb provides a means for measuring and controlling $f_{ceo}$ with a single detector \cite{Lind2020,Hoghooghi2022}.

In order to more accurately quantify the UV extent of the frequency comb, we additionally measured its spectrum using a Si-based array spectrometer in conjunction with a UV bandpass filter (max wavelength 400 nm). The result is show in Figure 4. Here we see that the generated spectrum extends down to approximately 330 nm.  We independently measured the average on-chip power of the UV-visible spectrum (350 nm - 550 nm) to be 1.5 mW with 9 mW (90 pJ) on-chip pump power. This implies an impressive 17\% conversion efficiency from the 1550 nm pump to the UV-visible spectrum. The conversion from 1550 nm to these shorter wavelengths can be controlled and enhanced by the linear chirp range and chirp rate. This is shown in Figure 4b, where we present our predictions of the integrated power of light spanning 350-550 nm with different poling conditions. For the parameters of the waveguides we have employed in the experiments, this simulation results shows the ideal efficiency can be near 22\%. Additional optimization of the waveguides can potentially yield efficiencies >30\% over specific wavelength regions, as detailed in the Supplement.

\section*{Discussion \& Conclusion}

Complex nonlinear interactions give rise to the broad bandwidth spectra across the visible and near UV. But as we discuss here, these interactions can be understood and subsequently tailored to optimize the supercontinuum generation. Different from the majority of other approaches, the devices we have fabricated rely most strongly on the $\chi^{(2)}$ nonlinearity and the engineered $\chi^{(3)}$ that arises from the cascading of the $\chi^{(2)}$ process. The  $\chi^{(2)}$ nonlinearity is orders of magnitude stronger than the intrinsic $\chi^{(3)}$ nonlinearity, and gives rise to efficient spectral broadening in the visible and near UV where phase-matching via $\chi^{(3)}$ is challenging due to the rapidly increasing normal dispersion.  

Figure 5 is a simulation that illustrates the evolution of the spectral broadening as function of propagation distance. The supercontinuum generation in the lithium niobate waveguide begins with engineered waveguide dispersion close to zero at 1550 nm. In this un-poled section of waveguide, self-phase modulation rapidly broadens the fundamental pulse at 1550 nm as shown in Figure 5c. Third harmonic generation is also seen in this section of the waveguide. Once the pulse enters the chirped poling section of the the waveguide, the second harmonic generation increases from lower frequency to high frequency, and is followed in similar fashion by fourth harmonic generation, as seen in Figure 5d-f. Finally, the spectrum in the region of the third harmonic that was initially produced in the un-poled region then grows by mixing of frequencies from the second harmonic and fundamental. Throughout the propagation in the poled region, we observe further spectral broadening across the entire visible and ultraviolet akin to effective $\chi^{(3)}$ via the cascading of the second order nonlinearity . This broadened spectrum is shown in Figure 5f.

In this work, we have demonstrated a multi-segement nanophotonic lithium niobate waveguide that combines both dispersion and quasi-phase matching engineering. This novel design leverages the interaction between $\chi^{(2)}$ and $\chi^{(3)}$ nonlinearities to generate continuous frequency comb spectra extending from 330 nm to 2400 nm with less than 100 pJ energy. This is accomplished with pumping at the wavelength of 1550 nm using mature Er:fiber laser technology.  
Furthermore, the coherence of the supercontinuum is verified by measuring $f_{ceo}$ from UV spectrum. 
Our results highlight the path to further enhance the interaction between $\chi^{(3)}$ and $\chi^{(2)}$ nonlinearities to enhance efficient visible and UV spectral generation. As discussed in the Supplement, optimized designs with additional high frequency poling in the first section of the waveguide can increase the conversion from 1550 nm into the UV with efficiency greater than 32\%.  Beside varying the length of the un-poled section, engineering the dispersion of waveguide can extend the bandwidth into the mid-infrared regime with intra-pulse or dual band different frequency generation. By engineering the phase (mis) matching in $\chi^{(2)}$ and $\chi^{(3)}$ effects in nanophotonic lithium niobate waveguides, it is now possible to generate a compact light source with coverage across the full transparency window of lithium niobate--from approximately 330 nm to 5000 nm.  Such design control will open new opportunities with the thin film lithium niobate platform for applications of spectroscopy and trace gas sensing, control of quantum systems, and atomic clocks.

\section*{Methods}

\subsection*{Waveguide Fabrication}
We use a commercially available X-cut wafer (NANOLN), with a \textasciitilde 700-nm-thick lithium niobate layer on top of a \textasciitilde 4.7-$\mu$m-thick silica buffer layer. We first deposit electrodes by e-beam evaporation (15 nm of Cr + 50 nm of Au) for periodic poling. We apply a \textasciitilde 700 V pulse over a 25 $\mu$m gap to produce the domain inversion. The poling period decreases linearly from $\Lambda=$12.5 µm to $\Lambda=$2.5 µm along a 3.6-mm length. Due to lateral domain growth during the poling process, the duty cycle for the metallic electrodes needs to be adjusted in order to obtain a 50\% duty-cycle along the poled length. Since this effect is more acute for shorter poling periods, we vary the electrodes duty-cycle linearly from 50\% to 20\% along the 3.6-mm poling length.

The waveguides are fabricated by first patterning a hydrogen silsesquioxane (HSQ) mask using e-beam lithography. The pattern is transferred to the lithium niobate layer by dry etching with Ar$^+$ plasma to a depth of 350 nm. Finally, the chip facets are polished by first coarse-lapping to reach a target of 3-mm unpoled waveguide length between the input facet and the poled area, followed by fine-polishing to improve the coupling efficiency.

\subsection*{Numerical Simulations}

The spectral evolution is modeled in the frequency domain with a single-mode unidirectional propagation equation\cite{Francois1991nonlinearFD, Kolesik2004UPE} that includes both second and third order nonlinearities.
\begin{equation}
    i \frac{\partial}{\partial z} a\!\left[\nu\right]
    = \left(\beta + i \frac{\alpha}{2}\right) a\!\left[\nu\right]
    + \frac{\omega \frac{\chi^{(2)}}{2}}{\sqrt{\epsilon_0 c^3 n^3 A}} \mathcal{F}\!\!\left[a\!\left[t\right]^2\right]\!\!\left[\nu\right]
    + \frac{\omega \frac{\chi^{(3)}}{2}}{\epsilon_0 c^2 n^2 A} \mathcal{F}\!\!\left[a\!\left[t\right]^3\right]\!\!\left[\nu\right]
\end{equation}
where $a\!\left[\nu\right]$ is the spectral amplitude, normalized to the pulse energy $e_p = \int{\left|a\right|^2 d\nu}$. The linear term includes the frequency-dependent angular wavenumber $\beta$ and gain/loss parameter $\alpha$. The nonlinear interactions are calculated in the time domain from powers of the carrier-resolved amplitude and then Fourier transformed ($\mathcal{F}$) into the frequency domain. Polling is implemented by changing the sign of $\chi^{(2)}$ at the boundary of each domain inversion. The strength of the nonlinear terms depend on the effective refractive index $n$ and mode area $A$~(1.095 µm$^{2}$), which were calculated using the mode solvers of Lumerical. The values of the refractive indices used in the mode solver are from Lumerical. The propagation loss was assumed to be 1.1 dB/cm and the values used for the nonlinear susceptibilities $\chi^{(2)}$ and $\chi^{(3)}$ were 30pm/V and 5200pm$^2$/V$^2$ respectively. For clarity, notation has been dropped for the frequency dependence of the waveguide parameters on the right-hand side of the equation. The model is integrated using an adaptive 3rd and 4th order embedded Runge-Kutta scheme in the interaction picture (ERK4(3)-IP)\cite{Balac2013ERK4(3)-IP}, and the full simulation framework is currently implemented in a fork of PyNLO\cite{pynlo}.





\section*{Code availability}
Simulations were carried out with the open-source code PyNLO\cite{pynlo}.

\bibliography{sample}

\section*{Acknowledgements} 
This work was supported by the National Science Foundation AST 2009982, EECS1846273 and QLCI Award No. OMA-2016244, the Jet Propulsion Laboratory, California Institute of Technology, under a contract with the National Aeronautics and Space Administration and funded through the internal Research and Technology Development program RSA 1671354, the Air Force Office of Scientific Research FA9550-20-1-0040, and the National Institute of Standards and Technology NIST on a Chip program. The device nanofabrication was performed at the Kavli Nanoscience Institute (KNI) at Caltech. The authors acknowledge helpful comments from Kristina Chang and Jennifer Black on the manuscript and valuable input from Stephanie Liefer in the early stages of this project.



\newpage
\section*{Figures \& Tables}

\begin{figure}[h]
\centering
    \includegraphics[width=16.5cm]{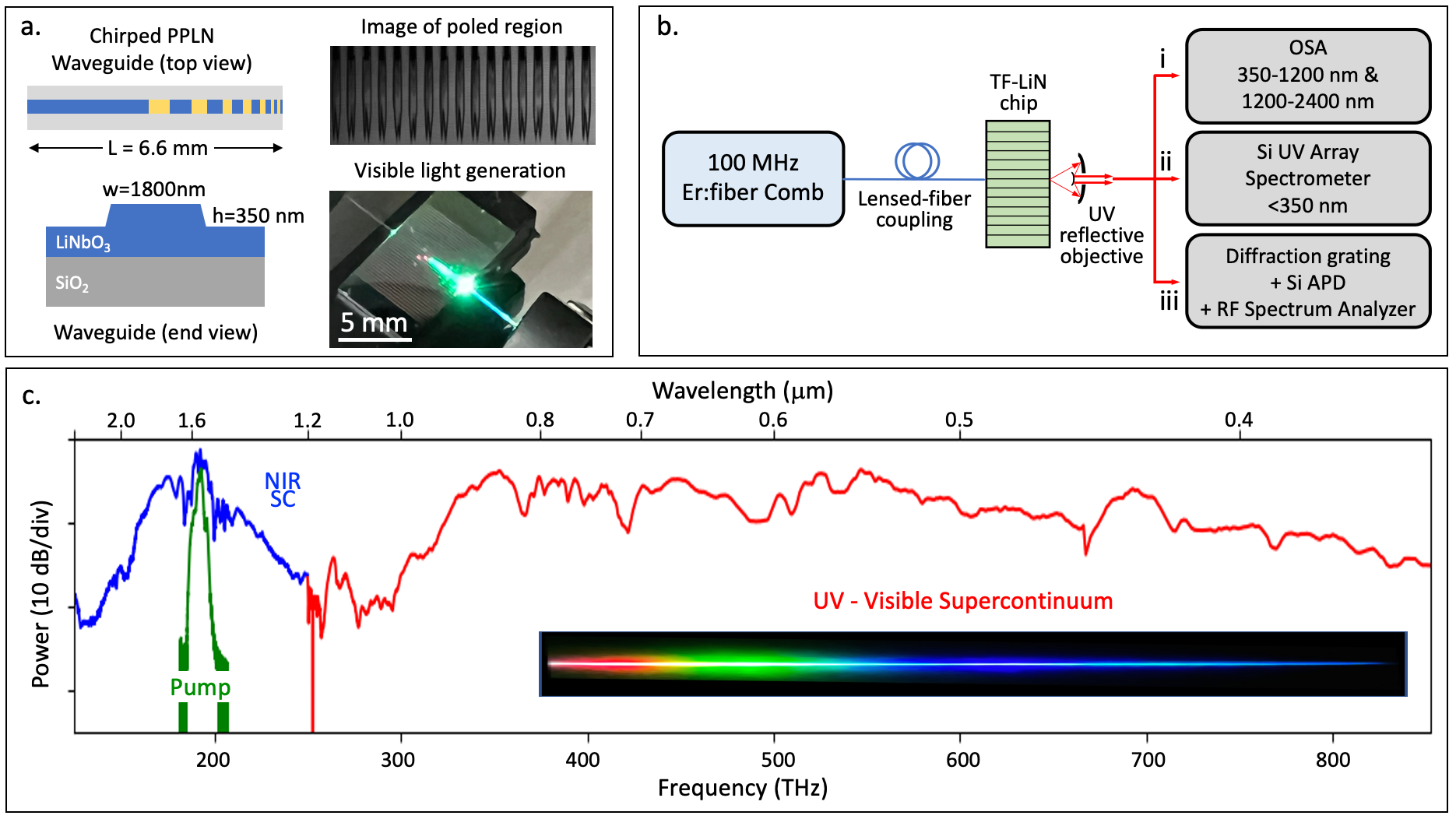}
 \caption{Nanophotonic lithium niobate waveguides for ultraviolet to near infrared frequency comb generation (a) Schematic of the waveguide with upoled region and then linear chirped poling with $\Lambda=12.5$ µm to 2.5 µm; Cross-sectional schematic of the waveguide with dimensions of w=1800 nm, h=350 nm.  The LN slab thickness is 360 nm.  Also shown is a photo of the waveguide generating white light and a two-photon microscope image of the shortest poling periods.  (b) Experimental setup for comb generation and characterization.  A suite of three different instruments are used to measure the spectrum and verify the optical coherence of the comb at wavelengths below 350 nm. (c) Supercontinuum generated with the nanophotonic waveguide.  The input spectrum of the Er:fiber laser at 1.55 µm is shown by the green line. Broadband frequency comb spectra generated in the LN waveguide with chirped poling is shown in blue and red. The insets show a photo of the dispersed visible spectrum of the comb.}
\end{figure} 

\newpage
\begin{figure}[h]
\centering
    \includegraphics[width=17cm]{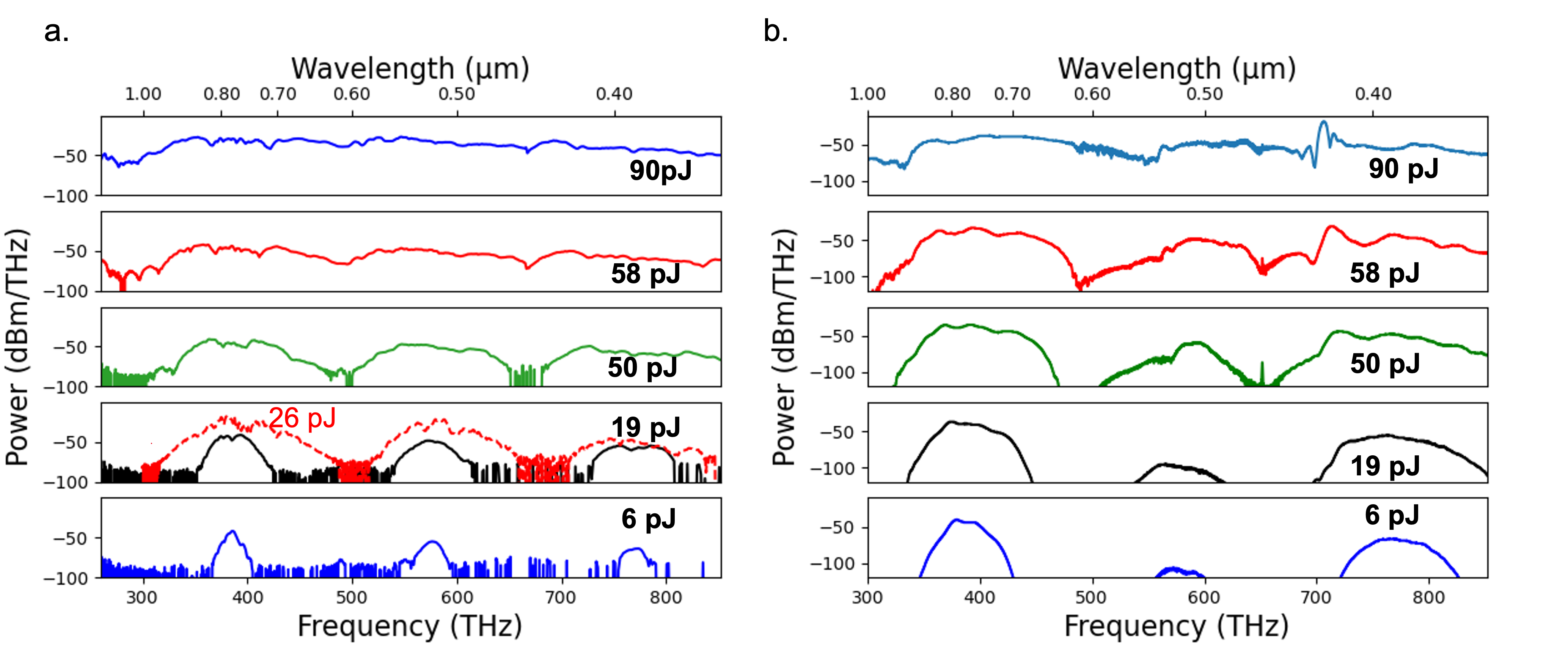}
 \caption{ Spectral evolution of broad bandwidth frequency comb generation. (a) Experimental results of the power spectral density with input driving pulse energies between 6 pJ and 90 pJ. All data were taken with a 100 MHz frequency comb pump soure, with the exception of the dashed red line at 26 pJ, which was obtained with a 10 GHz 1550 nm pump source. (b) Simulation results at the same pulse energies, showing good qualitative agreement with the experimental data. }
\end{figure} 

\newpage
\begin{figure}[h]
\centering
    \includegraphics[width=17cm]{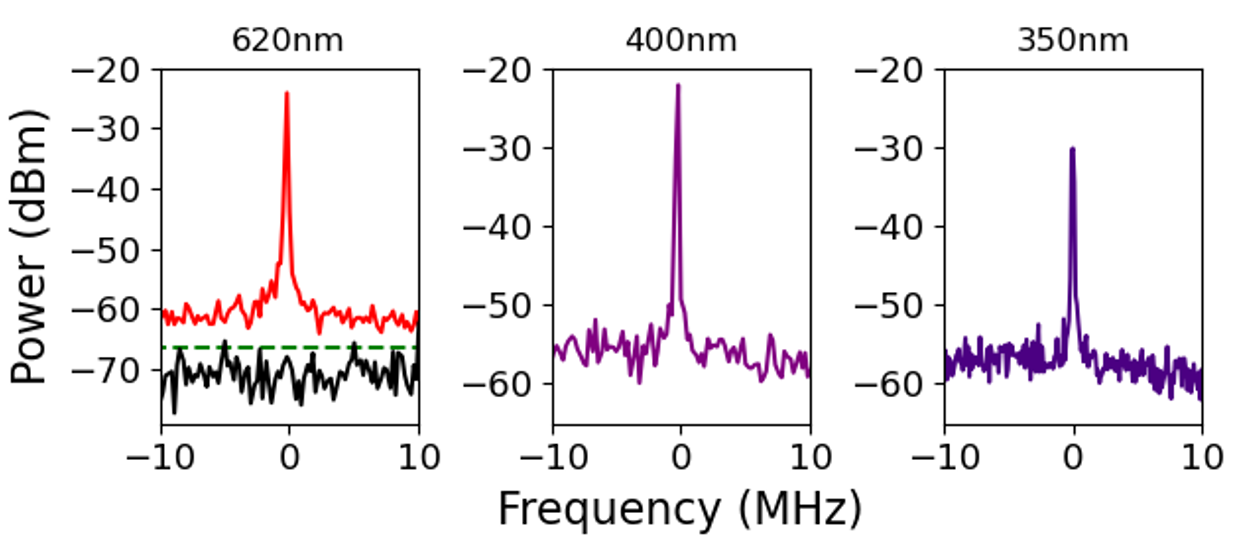}
 \caption{Broad bandwidth coherence and carrier-envelope offset frequency detection.  The $f_{ceo}$ beatnote is observed directly at wavelengths across the visible and UV. The SNR of $f_{ceo}$ at 620 nm and 400 nm are both greater than 35dB in 10 kHz resolution bandwidth. The SNR at 350 nm  is about 28dB. In the 620 nm plot, the green dashed line is the shot noise and black line is measurement noise floor.}
\end{figure} 

\newpage
\begin{figure}[h]
\centering
    \includegraphics[width=16cm]{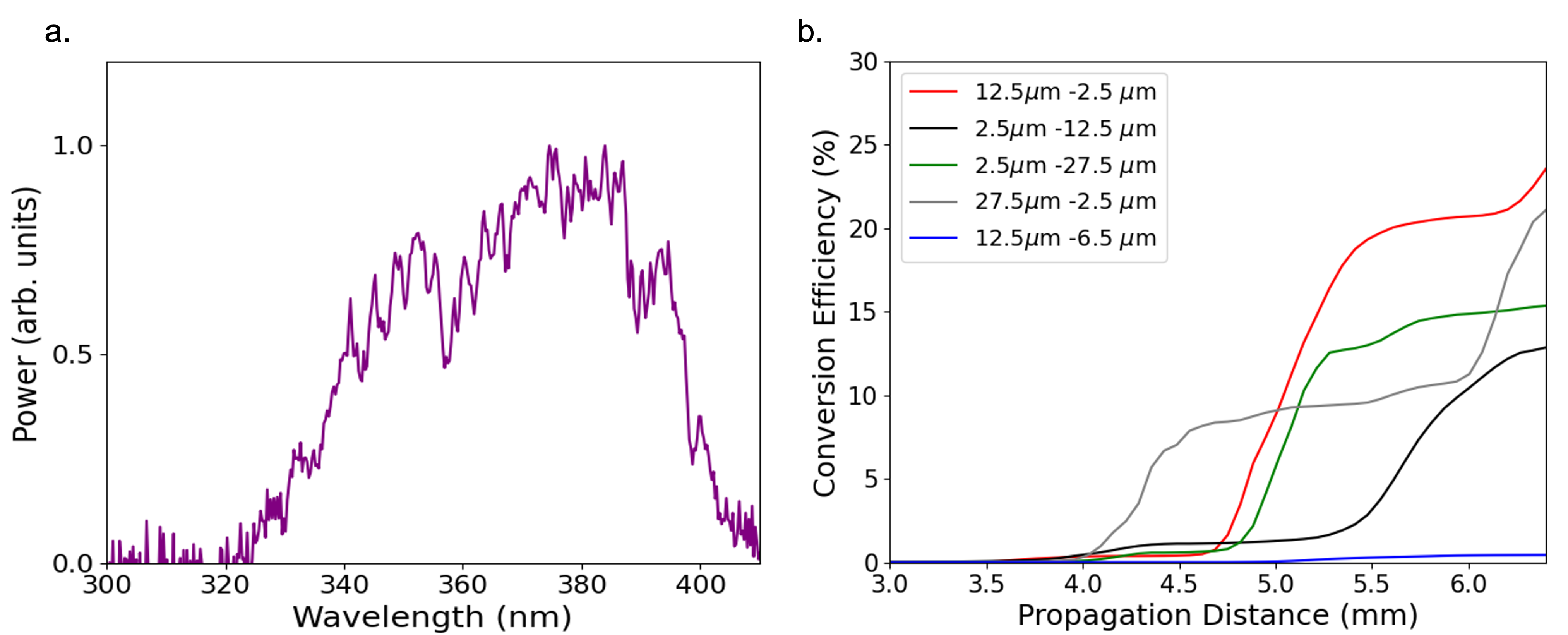}
 \caption{Ultraviolet spectral limits and efficiency (a) UV spectrum produced with the waveguide of Fig. 1a, and measured after a UV bandpass filter to eliminate any light at longer wavelengths. (b) A simulation showing the integrated power (350-550 nm) along the waveguide propagation direction. Each curve is from a different simulation in which the linear chirp rate and range were varied to illustrate the design capabilities.}
\end{figure} 

\newpage
\begin{figure}[h]
\centering
    \includegraphics[width=17cm]{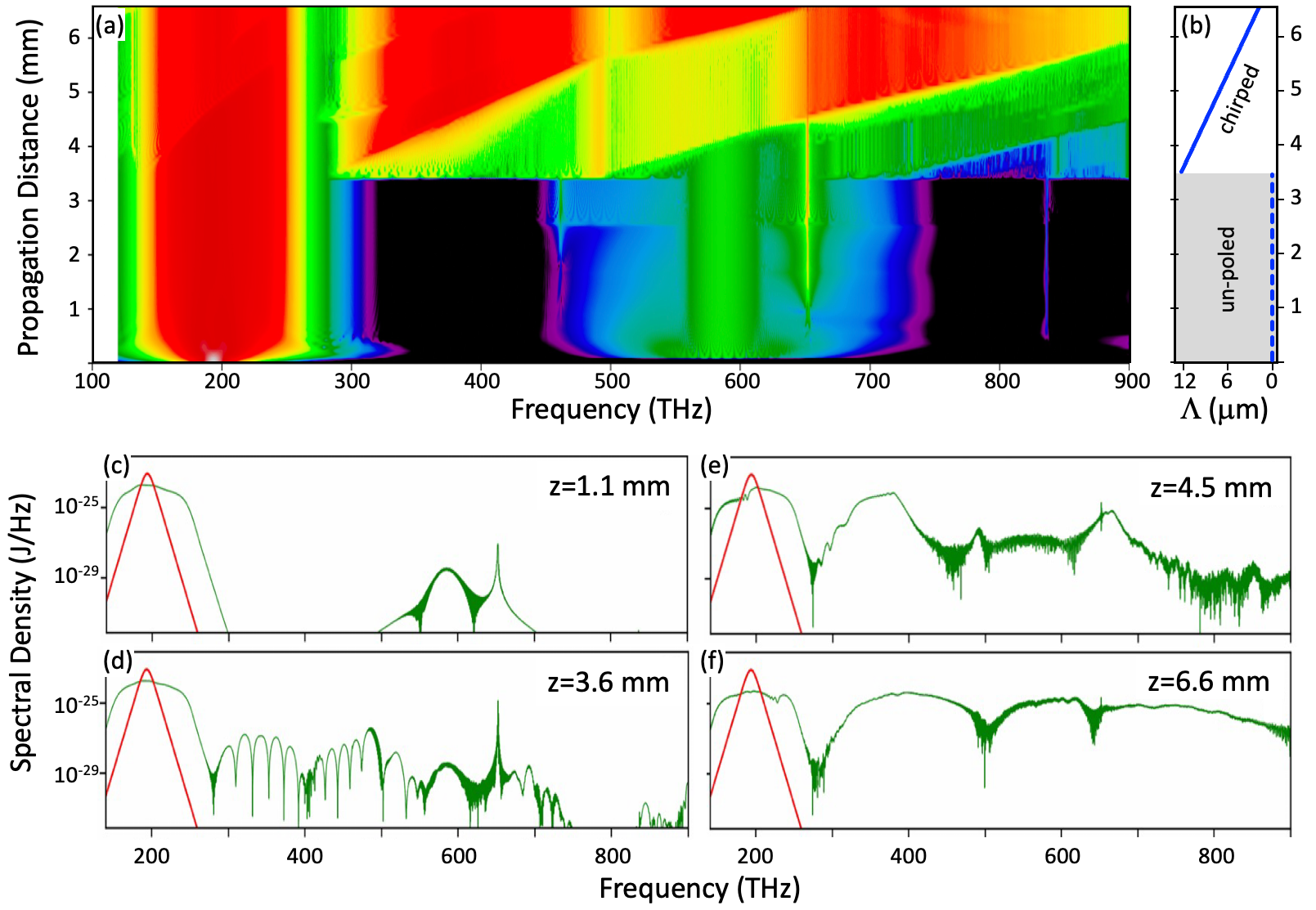}
 \caption{Nonlinear spectral evolution in thin-film lithium niobate waveguides with  $\chi^{(3)}$ and $\chi^{(2)}$ nonlinearities. (a) Simulation of the spectral evolution as function of pulse propagation distance.  (b) Map of the poling period ($\Lambda$) along the same propagation distance. The chirped poling starts at 3.6 mm and decreases linearly from $\Lambda=$12.5 µm to 2.5 µm. (c-f) Plots of the spectrum at  different propagation distances. The input spectrum for the simulation is shown in red. (c) The broadening of the input spectrum via the $\chi^{(3)}$ nonlinearity and non-phase-matched third-harmonic of the pump. (d) The spectrum at the beginning of the linearly chirped  poling region of the waveguide. (e-f) The visible spectrum is continually generated and spectrally broadened from the harmonic bands and the combination of $\chi^{(2)}$ effects.}
\end{figure}

\newpage

\section*{Supplementary Material}
\renewcommand\thefigure{\arabic{figure}}
\setcounter{figure}{0}
\vspace{0.5cm}   
{\textbf{Waveguide dispersion:}} In Supplemental Fig. 1, we present the calculated group velocity dispersion (GVD) for the waveguide used in this work, with h=350 nm.  In this figure, h is the etch depth into the 700 nm film and the waveguide width is w=1800 nm.  For comparison, a second  waveguide is also shown with h=410 nm.  For both waveguides in Fig. 1, the GVD is near zero at the pump wavelength of 1550 nm, and the GVD is strongly normal in the visible.  Below wavelengths of 1000 nm, the impact of geometric dispersion engineering is negligible for these waveguides.

\begin{figure}[h]
\centering
    \includegraphics[width=12.5cm]{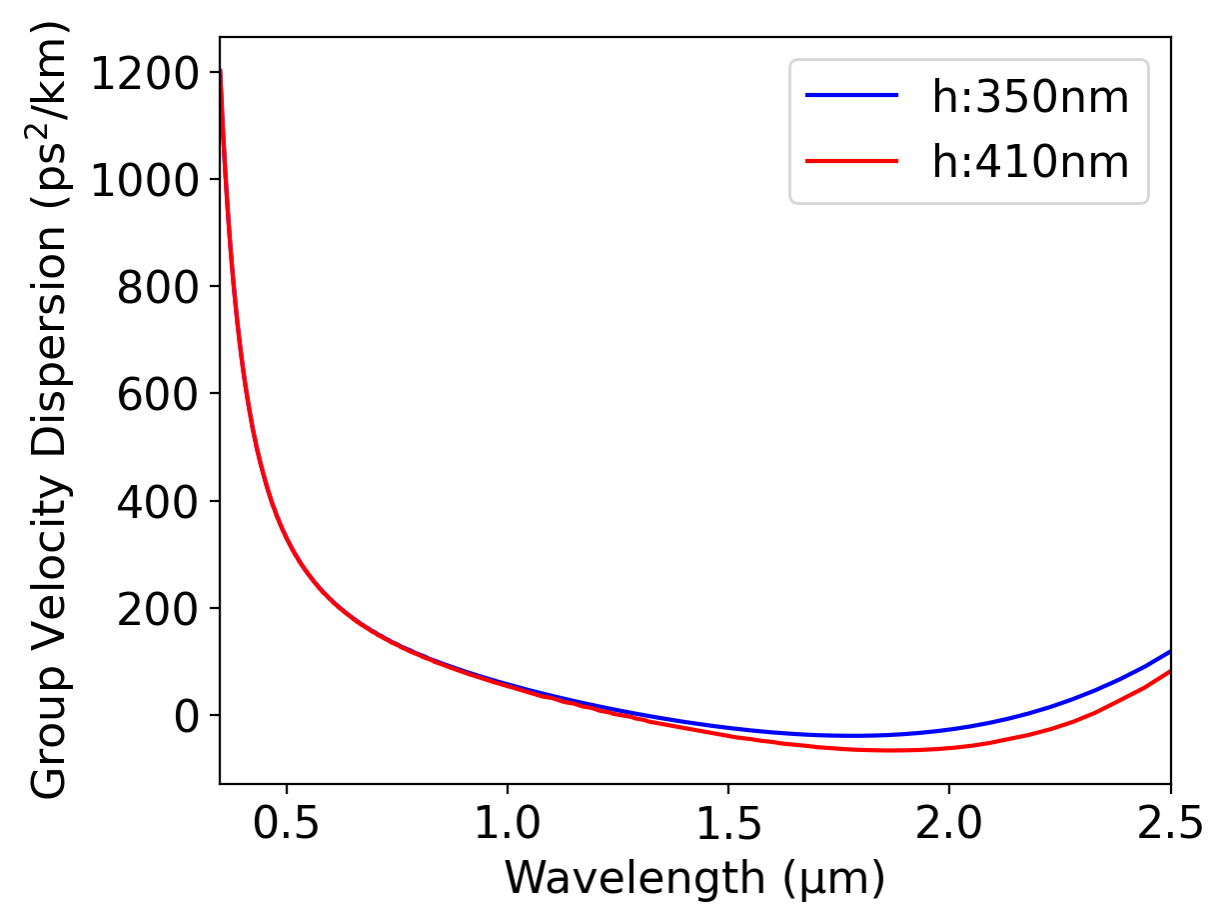}
 \caption{Calculated waveguide group velocity dispersion for the device used in this work.  The dispersion for two waveguides with width of w=1800 nm and etched height of h=350 nm and h=410 nm on a 700 nm film. }
\end{figure} 
  
{\noindent\textbf{Enhanced efficiency:}}
A design to enhance the power in the spectral window of 350-500nm is shown in Supplemental Fig. 2(a).  Here, we compare the design employed in the main text ("Design A") with a design that enhances the visible and UV generation ("Design B"). In the new design, we propose a normal dispersion LN waveguide with fixed short poling pattern over the first segment.  This period is chosen such that the 3rd and 4th harmonic spectra are enhanced and broadened with cascaded $\chi^{(2)}$ before the pulse enters the chirped poling segment. Figure 2(a) compares both designs, and "Design B" shows a theoretical improvement in power across 350-500 nm  by more than 4 orders of magnitude with only 30 pJ pump pulse energy. In this case, total conversion efficiency from 1550 nm to the 350-500 nm region is 15.9\%. 

In Fig. 5 of the main text we observe that the $\chi^{(3)}$ spectral broadening of the 1550 nm input pulse happens in the first few-hundred microns of propagation in the unpoled waveguide. Along with this spectral broadening, the pulse spreads in time and the peak power is decreased, which leads to lower efficiency in the nonlinear conversion. This observation leads us to investigate the impact of changing the length of the unpoled region, $L$, of the waveguide. (Note that $L=3$ mm for the actual device described in the main text).  Supplemental Fig 2(b) highlights the variation of conversion efficiency with $L$ from the 1550 nm pump into a window covering 350-490 nm. Here we observe that for $L=1$ mm a theoretical conversion efficiency of greater than 30\% is possible.  This type of analysis shows that further waveguide engineering can optimize the very-efficient projection of the 1550 nm light into specific spectral bands.  

\vspace{0.5cm}
{\noindent\textbf{Infrared generation:}}
Supplemental Fig. 2(c) shows a waveguide design that can lead to mid infrared (mid-IR) coverage. The height of the LN waveguide is 400 nm on a 710 nm film of LN on top of SiO$_2$ substrate. The length of waveguide is 6 mm, including a 5 mm unpoled segment, followed by a 1 mm segment with linearly chirped poling (3µm to 14.5µm). The waveguide dispersion is anomalous which enables soliton self-compression to increase the peak power and generate the desired bandwidth. The gap-free mid-IR coverage is generated with intrapulse difference frequency generation when the 50 pJ pump pulse propagates in the LN waveguide. Our model does not yet include the material absorption in the mid-IR in LN and thermal SiO$_2$ layers, but we see the potential to generate spectra across mid-IR with a sapphire substrate or fully air clad (suspended) waveguides. 

\begin{figure}[ht]
\centering
    \includegraphics[width=12.5cm]{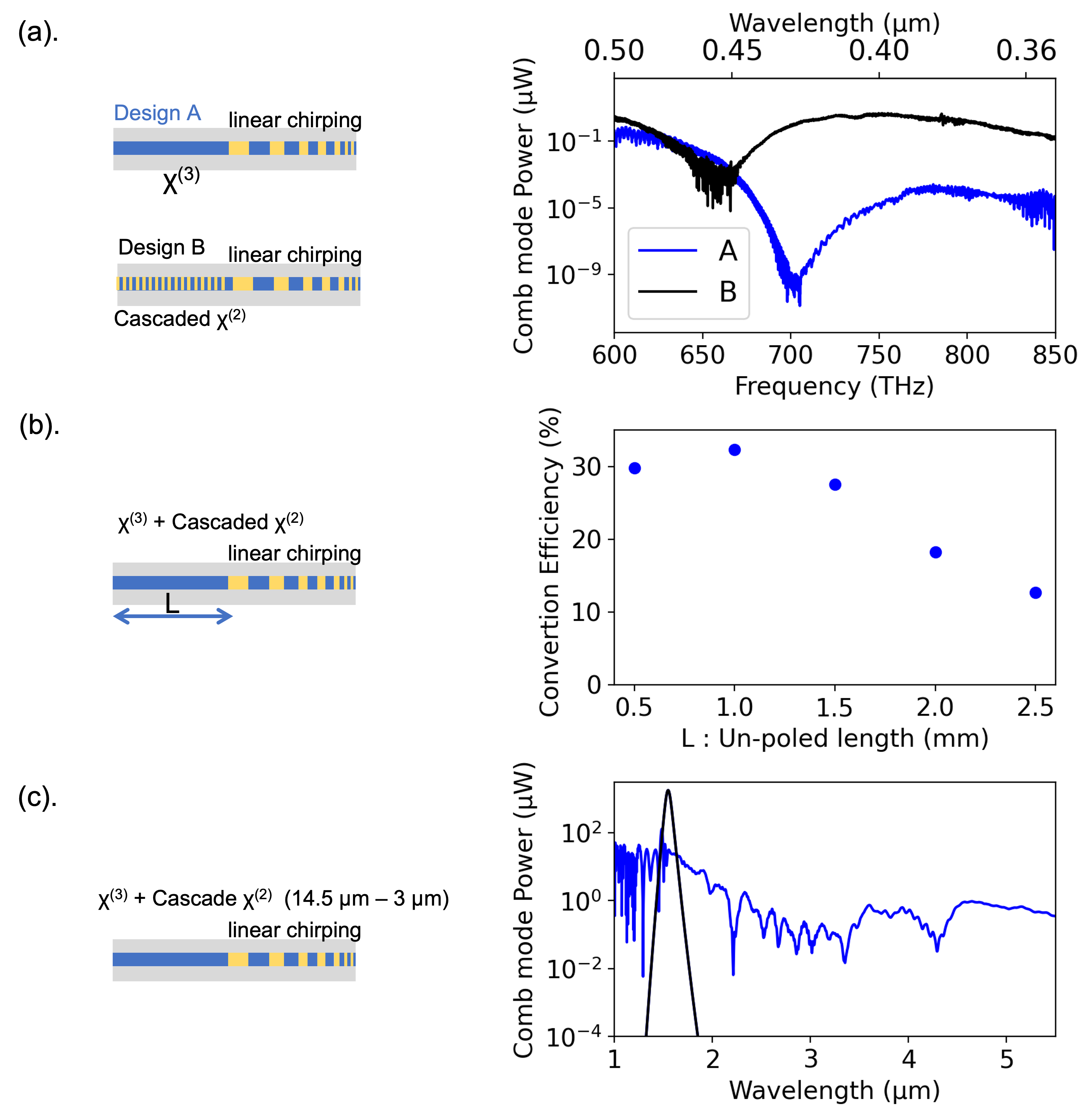}
 \caption{(a) Optimized wavelength coverage by design.  Given a fixed pump energy of 30 pJ, the wavelength coverage produced by “Design B” shows a theoretical improvement in power in the UV spectral region by more than 4 orders of magnitude when compared to the orignal “Design A”.(b) Impact of input unpoled waveguide length $L$ on the conversion efficiency from 1550 nm into the band spanning 350-490 nm. (c) Simulated anomalous dispersion waveguides with linear chirped poling pattern. The simulated spectra extends from 1 µm to beyond 5 µm (blue line) with a 5 mm long unpoled region that is followed by a  1 mm poled waveguide. The etch height is  $h=420$ nm and the width is $w=1200$ nm. The simulation assumes a a 50 fs and 50 pJ pump pulse (black line)}
\end{figure}

\end{document}